\documentclass[review]{elsarticle}

\usepackage{hyperref,graphicx,amsmath,booktabs,color,subfigure,siunitx}
\usepackage[version=4]{mhchem}		 	
\usepackage[makeroom]{cancel}
\biboptions{numbers,sort&compress}

\definecolor{darkRed}{RGB}{164,9,68}
\newcommand{\bi}{ \boldsymbol}

\begin{document}
\begin{frontmatter}

\title{Electromagnetically driven convection suitable for mass
  transfer enhancement in liquid metal batteries}

\author[hzdr]{Norbert Weber}
\author[hzdr]{Michael Nimtz}
\author[hzdr,turin]{Paolo Personnettaz}
\author[hzdr,monterrey]{Alejandro Salas}
\author[hzdr]{Tom Weier}

\address[hzdr]{Helmholtz-Zentrum Dresden -- Rossendorf, %
Bautzner Landstr.\ 400, Dresden, Germany}


\address[monterrey]{Instituto Tecnol\'ogico y de Estudios Superiores de Monterrey, Monterrey, Mexico}

\address[turin]{Politecnico di Torino, Corso Duca degli Abruzzi 24, 10129 Torino, Italy}

\begin{abstract}
Liquid metal batteries (LMBs) were recently proposed as cheap large
scale energy storage. Such devices are urgently required for 
balancing highly fluctuating renewable energy sources. During discharge,
intermetallic phases tend to form in the cathode of LMBs. 
These do not only limit the up-scalability, but also the 
efficiency of the cells. Generating a
mild fluid flow in the fully liquid cell will smoothen
concentration gradients and minimise the formation of
intermetallics. In this context we study electro-vortex flow
numerically. We simulate a recent LMB related experiment and discuss
how the feeding lines to the cell can be optimised to enhance mass
transfer. The Lorentz forces have to overcome the stable
thermal stratification in the cathode of the cell; we show that
thermal effects may reduce electro-vortex flow velocities
considerable. Finally, we study the influence of the Earth magnetic
field on the flow.
\end{abstract}


\end{frontmatter}

\clearpage

\section{Introduction}\label{intro}
Integrating highly fluctuating renewable energy sources (such as
photovoltaics and wind power) into the electric grid calls for large
scale energy storage. Such storage must be, first of all, safe and
cheap. The liquid metal battery (LMB) promises both. After being
intensively investigated in the 1960s, and abandoned later, LMB
research experienced a renaissance some ten years ago. For an overview
of the pioneering work, see \cite{Cairns1967,Chum1980,Chum1981}
(recommended \cite{Swinkels1971}) and for the recent work
\cite{Kim2013b} and \cite{Kelley2017}.

\begin{figure}[bth]
\centering
\includegraphics[height=4.8cm]{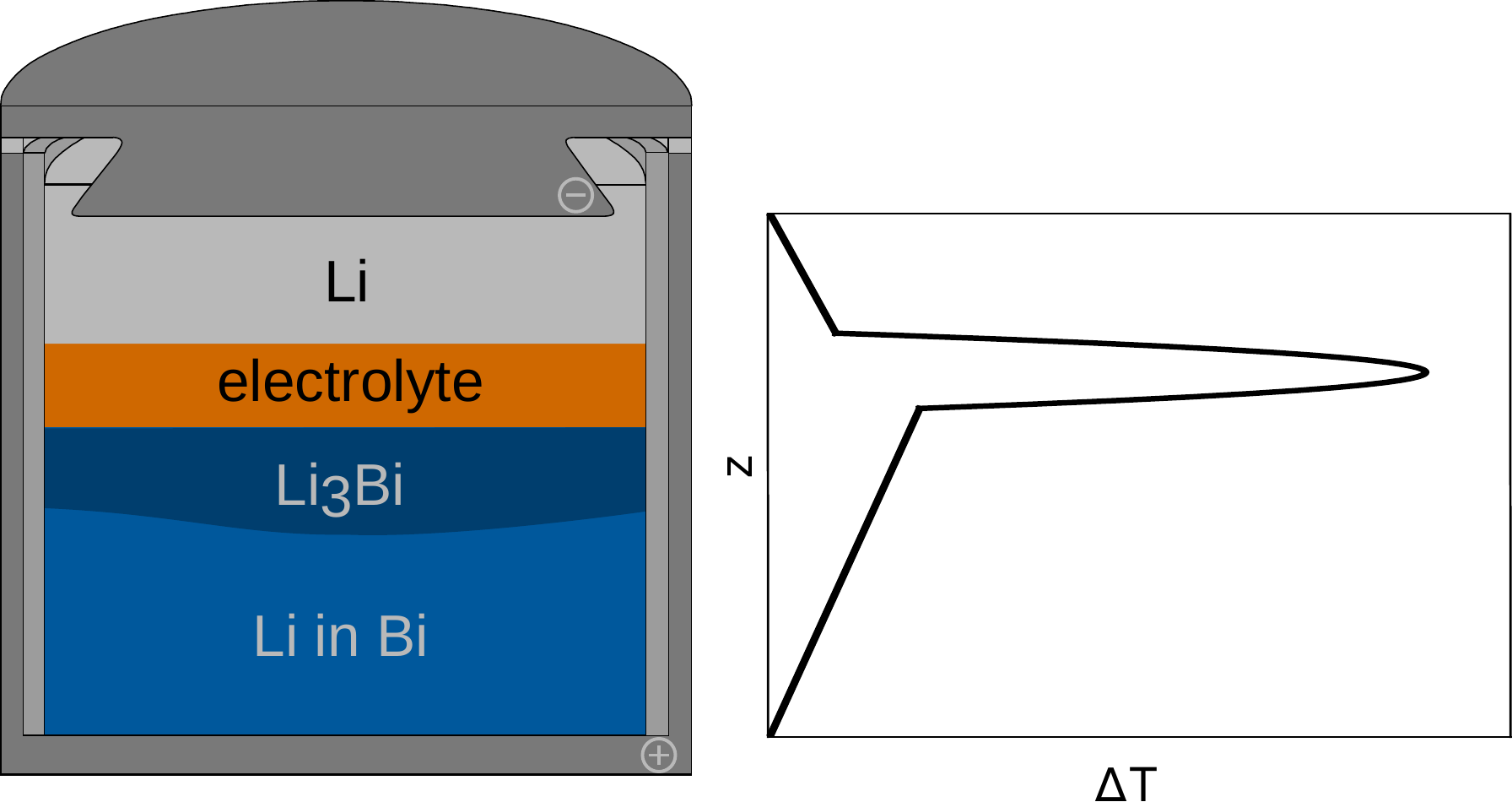}
\caption{Sketch of a typical Li$||$Bi liquid metal battery with an
  intermetallic phase forming in the cathode (left) and vertical
  temperature distribution in the three layers for pure diffusion (right).}
\label{f1}
\end{figure}
Fig.~\ref{f1}a shows a sketch of a typical LMB. A dense metal on the bottom
(cathode, positive electrode) is separated by a liquid salt from a
lighter metal at the top (anode, negative electrode). All three phases
float above each other; the salt acts as the electrolyte. The word ``liquid
metal battery'' names only a family of electrochemical cells (which may consists of
many different active metals combinations). Typical cell couples include Ca$||$Bi
\cite{Kim2013a,Ouchi2016}, Ca$||$Pb \cite{Poizeau2012},
K$||$Hg \cite{Agruss1962,Agruss1967}, Li$||$Bi
\cite{Lawroski1963,Foster1964,Vogel1966,Cairns1967,Shimotake1969,Ning2015},
Li$||$Pb
\cite{Cairns1967,Wang2014}, Li$||$Sb \cite{Kim2013b,Wang2014},
Li$||$Sn
\cite{Vogel1965,Vogel1966,Foster1966b,Cairns1967,Hesson1967},
Li$||$Zn \cite{Cairns1967}, Mg$||$Sb
\cite{Bradwell2011,Bradwell2012,Kim2013b}, Na$||$Bi
\cite{Foster1967a,Cairns1967,Hesson1967,Vogel1967,Vogel1966,Vogel1965b,Vogel1965,Shimotake1967,Vogel1966b},
Na$||$Hg \cite{Heredy1967,Kim2013b,Spatocco2014}, Na$||$Pb
\cite{Cairns1967,Hesson1967,Vogel1966,Kisza1995,Vogel1966b},
Na$||$Sn
\cite{Weaver1962,Agruss1963,Agruss1967,Cairns1967,Hesson1967,Vogel1965}
and Na$||$Zn \cite{Xu2016,Xu2017} as well as exotic ones such as
Li$||$Se
\cite{Cairns1967,Cairns1969b,Cairns1969c} or Li$||$Te                                                                                                
\cite{Vogel1966,Cairns1967,Cairns1969b,Cairns1969c,Shimotake1969}.

During discharge,
the anode metal is oxidised, crosses the electrolyte layer and alloys
in the bottom layer with the dense metal
(``concentration cell''). Commonly, the ohmic resistance of the
electrolyte layer represents the most important overvoltage. However,
at higher discharge currents concentration polarisation enters the
field
\cite{Agruss1963,Heredy1967,Agruss1967,Foster1967b,Bradwell2012,Kim2013b}.
Example: when \emph{discharging} a Li$||$Bi cell, Li-rich alloy will concentrate at the
cathode-electrolyte interface. When a certain local concentration is
exceeded, a solid intermetallic phase (Li$_3$Bi) will form (fig.~\ref{f1}a)
\cite{Cairns1967,Vogel1967}. Such intermetallics often float on the cathode
metal \cite{Ouchi2014}. Sometimes they expand during
solidification. As the walls impede a lateral expansion, the
intermetallic will form a dome until finally short-circuiting the
electrolyte. Especially in Ca based cells, locally growing dendrites
may additionally short-circuit the cell \cite{Kim2013a}. Besides of
all the mentioned drawbacks, the formation of intermetallics has one
advantage: it removes anode metal from the melt and keeps thereby the
voltage constant. It should be also mentioned that some intermetallics
have high electrical resistances while others are good conductors.

When \emph{charging} the cell of fig.~\ref{f1}a, the positive electrode-electrolyte
interface will deplete of Li and a similar concentration gradient may
develop \cite{Vogel1967}. This effect is undesirable, too. Finally,
all the same effects may theoretically happen in the anode
compartment, too, if an alloyed top electrode is used (e.g. Ca-Mg
\cite{Bradwell2011,Ouchi2016}). However, such effects were not
reported, yet.

It was early proposed that a mild fluid flow may counterbalance
concentration gradients and increase thereby the efficiency of LMBs
\cite{Foster1967b,Cairns1967,Vogel1967}. While ``mechanical stirring''
\cite{Foster1967b,Cairns1967} seems difficult to realise, a localised
heating or cooling inducing thermal convection may be a very good
option \cite{Bradwell2011a,Bradwell2015}. Electro-vortex flow (EVF)
may be used for an efficient mass transfer enhancement,
too \cite{Weber2014b,Stefani2015,Ashour2017a}. Simply saying, EVF
always may develop when current lines are not in parallel. It can therefore easily be
adjusted by choosing the diameter/geometry of the current collectors
and feeding lines appropriately. EVF drives a jet away from the wall,
forming a poloidal flow \cite{Lundquist1969}. For a classical example 
of the origin of EVF, see Lundquist \cite{Lundquist1969} and Shercliff
\cite{Shercliff1970}, for a good 
introduction Davidson \cite{Davidson2001} and a detailed overview
including many experiments Bojarevics et al. \cite{Bojarevics1989}. The
relevance of EVF for LMBs is outlined by Ashour et al. \cite{Ashour2017a}. It should
also be mentioned that other flow phenomena like the Tayler
instability
\cite{Vandakurov1972,Stefani2011,Seilmayer2012,Weber2013,Weber2014,Weber2015b,Herreman2015,Stefani2017,Weier2017}, 
Rayleigh-B\'enard convection \cite{Shen2015,Koellner2017} or interface
instabilities
\cite{Zikanov2015,Weber2017,Weber2017a,Bojarevics2017,Horstmann2017}
may enhance mass transfer in LMBs, as well.

This article is dedicated (mainly) to electro-vortex flow. It's
aim is twofold: first, we will show how the connection 
of the supply lines to the cell influences the flow. Second, we study
how electro-vortex flow and stable thermal 
stratification interact. For this
purpose we combine numerical simulation with a simple 1D heat conduction
model. These models -- and the experiment which inspired our studies
-- are described in the following section.

\section{Physical, mathematical and numerical model}\label{s:model}
In this section we will first present the experiment \cite{Kelley2014}
which inspired this article. Thereafter we explain the way in which we
estimate the temperature gradient appearing in the cathode of a liquid
metal battery (LMB). Finally, we give an introduction to the 3D
numerical models used.

\subsection{Liquid metal electrode experiment}
\begin{figure}[tbh]
\centering
  \includegraphics[width=0.7\textwidth]{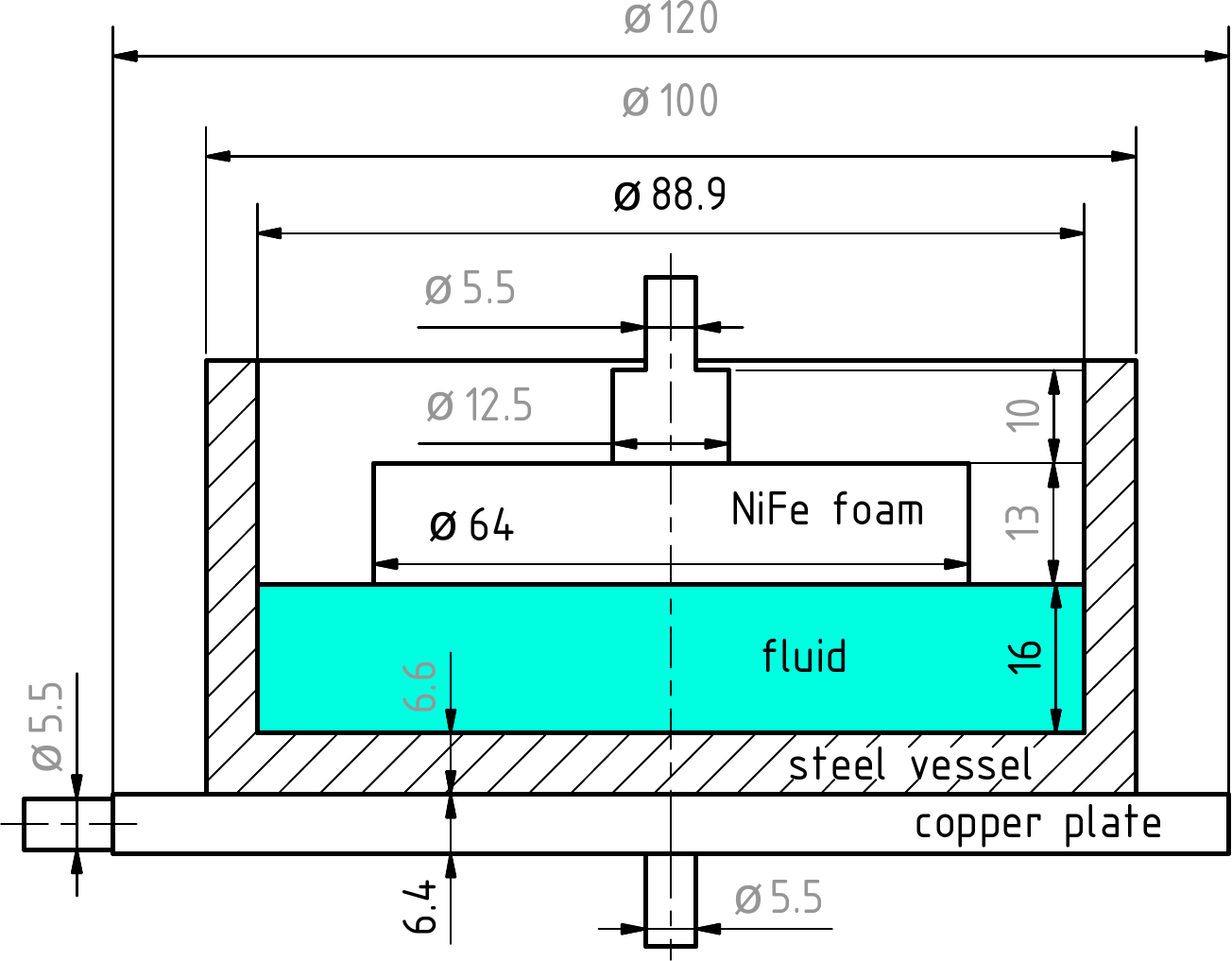}
\caption{Dimensions of the experiment and simulation model (in mm). The gray values
are not exactly known; they are estimated from the sketch in \cite{Kelley2014}. 
The wires are assumed to be made of copper.}
\label{f2}
\end{figure}
Fig.~\ref{f2} illustrates the mentioned experiment, conducted by Kelley
\& Sadoway \cite{Kelley2014}. A cylindrical steel vessel
contained a melt of eutectic lead-bismuth.
An electric current (up to \SI{0.375}{\ampere/\centi\meter\square}) was applied between a bottom and top electrode. The
bottom current was supplied centrically or laterally. The upper
electrode consisted of a nickel-iron foam; such foam is often used in
LMBs to contain the anode metal \cite{Kelley2017}. As the setup was heated from
below, Rayleigh-B\'enard cells appear. If an internal current was
applied, the flow became much more regular at \SI{0.05}{\ampere/\centi\meter\square}. It was
deduced by the authors that convection cells align with the magnetic field. We will
demonstrate how electro-vortex flow may give an alternative
explanation for the increase in order.

We use the following material properties of  lead bismuth eutectic (LBE) at
\SI{160}{\celsius} \cite{Ashour2017a}:
a kinematic viscosity of
$\nu=2.7\cdot10^{-7}$\,m$^2$/s, a thermal expansion coefficient of
$\beta = 1.3\cdot 10^{-4}$\,K$^{-1}$, an electrical conductivity of
$\sigma=9\cdot 10^5$\,S/m, a density of $\rho = 10\,505$\,kg/m$^3$, a
specific heat capacity of $c_p = 148$\,J/kg/K, a thermal conductivity of
$\lambda = 10$\,W/m/K, a thermal diffusivity of $\alpha=6\cdot
10^{-6}$\,m$^2$/s, a Prandtl number of $Pr=0.04$ and a sound velocity
of $u_\text{s} = 1\,765$\,m/s \cite{Sobolev2007,Sobolev2010,NEA2015}. The
electrical conductivity of the vessel is assumed to be $\sigma=1.37\cdot
10^6$\,S/m and of the wires and copper plate $\sigma = 58.1\cdot
10^7$\,S/m. The electrical conductivity of the Fe-Ni foam is not easy to
determine \cite{Dharmasena2002,Huang2009}, especially because it is
not sure if the liquid metal enteres the pores. We do not model the
porosity and use an electric conductivity of $\sigma=1.37\cdot 10^6$\,S/m 
without further justification.


\begin{figure}[tbh]
\centering
  \includegraphics[width=0.9\textwidth]{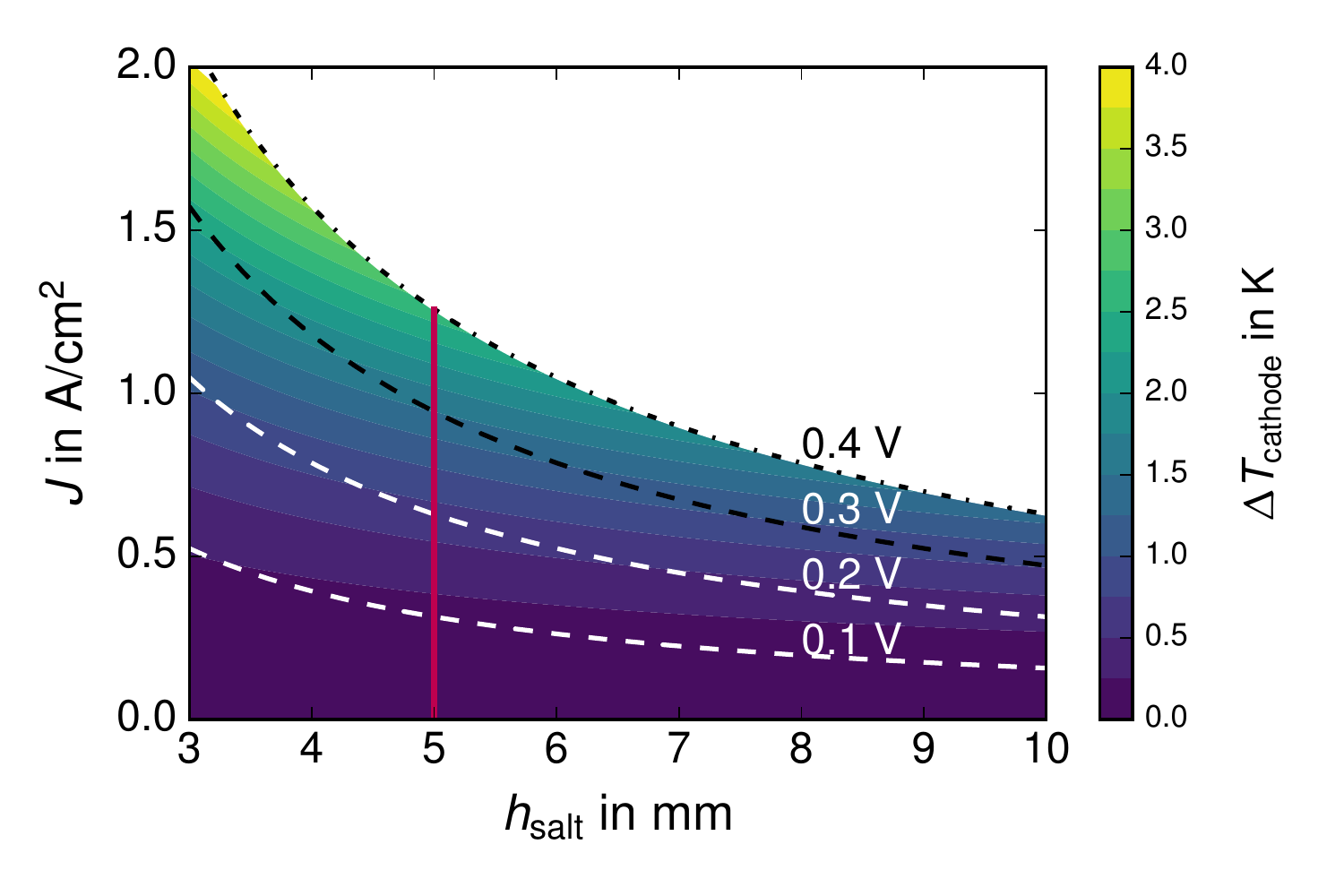}
\caption{Voltage drop and temperature difference in the cathode 
for pure conduction in a Li$||$Bi cell \cite{Personnettaz2017}.}
\label{f3}
\end{figure}

Geometrically, the described experiment perfectly represents a liquid
cathode of an LMB. However, the
temperature gradient in a working LMB depends on the boundary
conditions. For a single cell with insulated lateral walls it
will rather be opposite to that
in the experiment. As the electrolyte layer has the highest electrical resistance
(four orders larger than the metals), most heat will be generated 
there \cite{Shen2015}. Fig.~\ref{f1}b shows a typical vertical temperature profile
through all three layers. If no thermal management system
induces additional temperature gradients (as suggested in 
\cite{Bradwell2011,Bradwell2015}) a stable thermal stratification is
expected in the cathode. We will study here, if electro-vortex
flow can overcome this stratification.
For this purpose, we need a characteristic vertical temperature
gradient. As no LMB exists, which operates at such low temperature as
the experiment, we use a similar cell to define a typical vertical
temperature gradient: a Li$||$Bi LMB operating at \SI{450}{\celsius}. 

\subsection{Heat conduction model}
The temperature difference between top and bottom of the cathode of an LMB can
be estimated using the simple 1D heat conduction model developed by
Personnettaz \cite{Personnettaz2017} (for 3D studies of heat
transfer in Li-LMBs, see \cite{Koellner2017,Wang2015}). 

The model was developed with the
assumption of a fluid at rest, constant and homogeneous material
properties and a uniform current density distribution. The heat generation was
included only in the form of
Joule heating in the electrolyte layer, due to its high electrical resistivity ($\frac{\rho_\text{el,salt}}{\rho_\text{el,metal}}>10^3$).
The lateral wall of the cell was considered adiabatic and the top and the bottom boundaries
are set to a constant temperature $T = \SI{450}{\celsius}$.
Thanks to the mentioned hypotheses a 1D model along the vertical coordinate
is able to completely describe the temperature distribution inside the cell.
This profile provides an upper bound for the temperature and a valuable initial guess
of the temperature stratification in the cathode.
The cell studied by Personnettaz was a Li$|$KCl-LiCl$|$Bi LMB \cite{Personnettaz2017}.
We use his model to estimate
the temperature difference between the top and bottom of the cathode as 
\begin{eqnarray}
\Delta T=\frac{h_\text{Bi} h_\text{salt}   \left(2 h_\text{Li} \lambda_\text{salt} + h_\text{salt} \lambda_\text{Li}\right)\rho_\text{el,salt}J^2}{2 h_\text{Bi} \lambda_\text{Li} \lambda_\text{salt} + 2 h_\text{Li} \lambda_\text{Bi} \lambda_\text{salt} + 2 h_\text{salt} \lambda_\text{Bi} \lambda_\text{Li}}
 \end{eqnarray}
with $h$, $\lambda$, $\rho_\text{el,salt}$ and $J$  denoting the layer heights,
the thermal conductivities, the specific resistance of the salt and the current density.
The thickness of the cathode is set to \SI{16}{\milli\meter} as in
the experiment and the Li-layer to \SI{32}{\milli\meter}
in order to maximise the cell
capacity (see \cite{Personnettaz2017}).
The geometrical parameters and the transport properties
employed are collected in tab. \ref{tab:properties}.
Depending on the current density $J$ and
thickness of the electrolyte $h_\text{salt}$, $\Delta T$
over the cathode changes as illustrated in fig.~\ref{f3}.

We limit the maximum ohmic over-voltage to \SI{0.4}{\volt} which 
corresponds to a cell efficiency of about \SI{66}{\percent}.
We assume further that the electrolyte  is \SI{5}{\milli\meter}
thick (realistic values are 3-\SI{15}{\milli\meter} \cite{Weber2016a}) and
find the following law that describes the dependence of
the temperature difference by the current $I$ (assuming a cathode base area of
\SI{62.1}{\centi\meter\squared}) as
\begin{eqnarray}
 \Delta T (\text{K}) = \SI{4.37e-4}{(\kelvin/\ampere^2)}\cdot I^2(\si{\ampere^2}).
\end{eqnarray}
This formula provides only a rough estimate of a possible temperature
difference in the cathode. The multiplicative factor strongly depends on
the materials selection and their transport properties
at the operation temperature.
Anyway it provides a first estimate that allows to study
the competition of two current driven phenomena, electro-vortex flow
and thermal stratification, in a low temperature liquid metal experiment.
\begin{table}[h!]\label{t:conduction}
\centering
\caption{Properties of the pure substances at $T=\SI{450}{\celsius}$ and dimension for the pure heat conduction model
\cite{Ohse1985,Janz1979a,Janz1988,Fazio2015,Cornwell1971}. The molten salt employed is \ce{KCl-LiCl}.}\label{tab:properties}
\begin{tabular}{llrrr}
\hline
\multicolumn{1}{c}{property} & \multicolumn{1}{c}{unit} & \multicolumn{1}{c}{Li} & \multicolumn{1}{c}{salt} & \multicolumn{1}{c}{Bi} \\
\midrule
$h_i$ & \si{\milli\meter} & 32 & 5 & 16 \\ \hline
$k$   & \si{\watt/(\meter.\kelvin)}& 53.0 & 0.69 & 14.2 \\
$\sigma_\text{el}$& \si{\siemens/\meter}& -- & 157 & -- \\
$\rho_\text{el}$  & \SI{e-2}{\ohm\meter}& -- & 6.36 &  --\\
\hline
\end{tabular}\vspace{3pt}\end{table}

\subsection{Numerical model}
The numerical model is implemented in OpenFOAM \cite{Weller1998}; for all
details and the validation of the
electro-vortex flow solver see
\cite{Weber2017b}. Basically, it computes the electric potential $\phi$
and current density $\bi J$ on a global mesh as
\begin{eqnarray}
  \nabla\cdot\sigma\nabla\phi &=& 0\\
  \bi J &=& -\sigma\nabla\phi
\end{eqnarray}
with $\sigma$ denoting the electrical conductivity. All conducting
regions (of different conductivities -- see fig. \ref{f3}) are fully coupled. The results are
then mapped on a separate fluid mesh (blue area in fig. \ref{f3}). Induced currents and magnetic
fields are neglected, which is justified as long as the velocities are
small. On the fluid mesh the following set of equations is solved:
\begin{eqnarray}
&&\frac{\partial \bi u}{\partial t} + \left( \bi u \cdot \nabla \right)
\bi u = - \nabla p + \nu \nabla^2 \bi u + \frac{\bi J\times\bi
  B}{\rho}\\
&&\bi B(\bi r) = \frac{\mu_0}{4\pi}\int \frac{\bi J(\bi
  r')\times(\bi r
  - \bi r')}{|\bi r - \bi r'|^3}dV'
\end{eqnarray}
with $t$, $\bi u$, $p$, $\nu$, $\rho$, $\mu_0$, $\bi r$ and $dV$ denoting the
time, the velocity, the pressure, the kinematic viscosity, the density, the vacuum
permeability, the coordinate and the differential volume, respectively.
The fluid mesh has at least 200 cells on the
diameter, which is fine enough according to \cite{Ashour2017a}.

Thermal effects are modelled in the fluid only using 
the Oberbeck-Boussinesq
approximation \cite{Oberbeck1879} (for its validity, see
\cite{Gray1976,Ashour2017a}). The following set of equations is solved
\begin{eqnarray}
  \frac{\partial \bi u}{\partial t} + \nabla\cdot(\bi u\bi u)
  &=&-\nabla p_d + \nu\nabla^2\bi u - \bi g\cdot\bi r\nabla\rho_k +
    \frac{\bi J\times\bi B}{\rho_0}\\
  \nabla\cdot\bi u &=& 0\\
\frac{\partial T}{\partial t} + \nabla\cdot(\bi u T) &=&
\frac{\lambda}{\rho_0 c_p}\nabla^2 T 
\end{eqnarray}
with $\bi u$, $p$, $\nu$, $\bi g$, $\bi r$, $T$, $c_p$ $\bi
J$ and $\sigma$ denoting velocity, pressure, kinematic viscosity,
gravity, position vector, temperature, specific heat capacity, current
density and electrical conductivity, respectively. The density
$\rho=\rho_0\rho_k = \rho_0(1-\beta(T-T_{\text{ref}}))$ is calculated using
the mean density $\rho_0$ at reference temperature $T_{\text{ref}}$ and the
coefficient of thermal expansion $\beta$. $\bi J$ and $\bi B$ are
determined by the electro-vortex solver as described above; the resulting 
Lorenz force is assumed to be constant in time, i.e. induced currents
are neglected. At least 250 cells on the diameter and strongly refined 
boundary layers are used. No-slip boundary conditions are used for velocity.
This is justified even for the free surface due to the oxide film 
formed there \cite{Cramer2014}. The side walls are modelled as adiabatic
while a constant temperature is applied at the top and bottom boundaries.

\section{Results}
\begin{figure}[tbh]
\centering
\includegraphics[width=\textwidth]{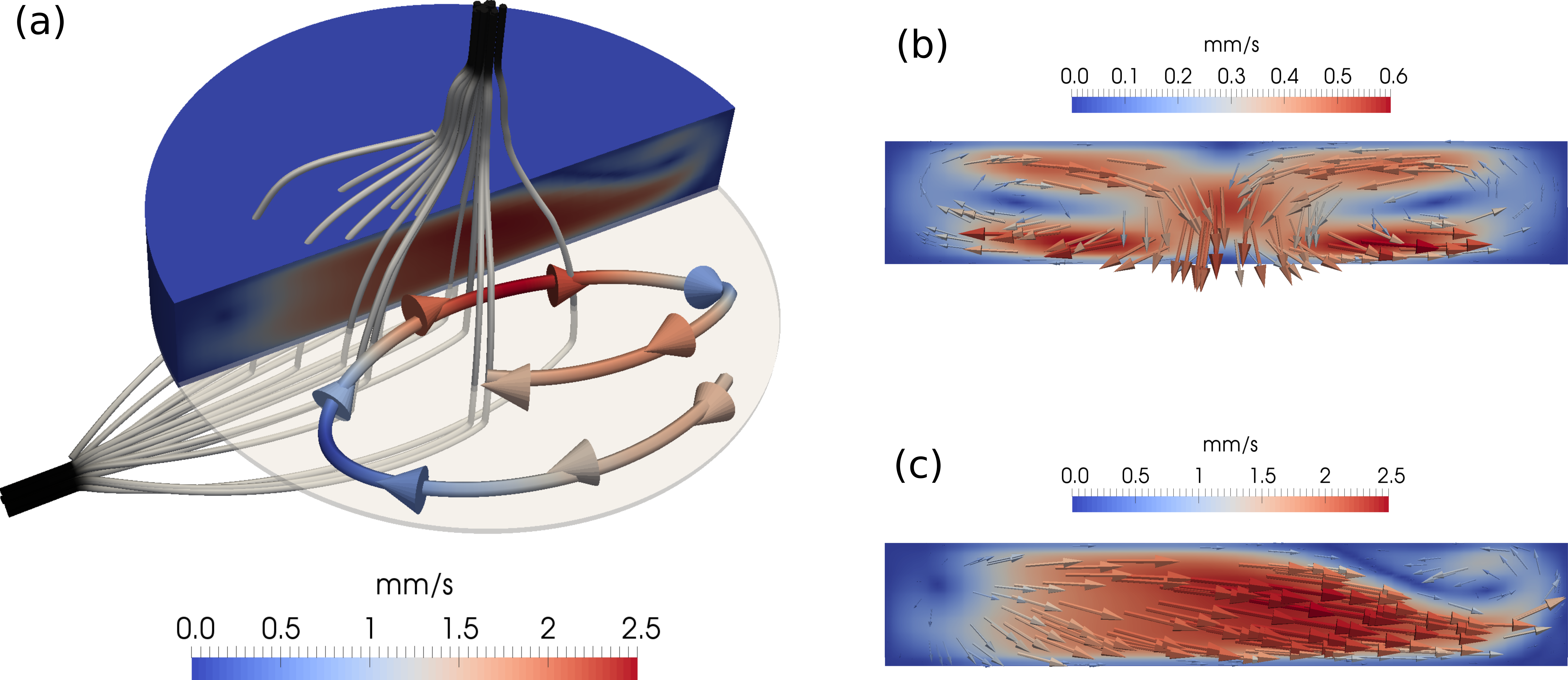}
\caption{Current path and velocity streamlines for a current supply
  from the side (a). Velocity on a vertical plane for symmetric (b)
and lateral current supply (c). The current is $I=40$\,A; the results
show electro-vortex flow alone.}
\label{f4}
\end{figure}

\begin{figure}[hbt!]
\centering
\includegraphics[width=\textwidth]{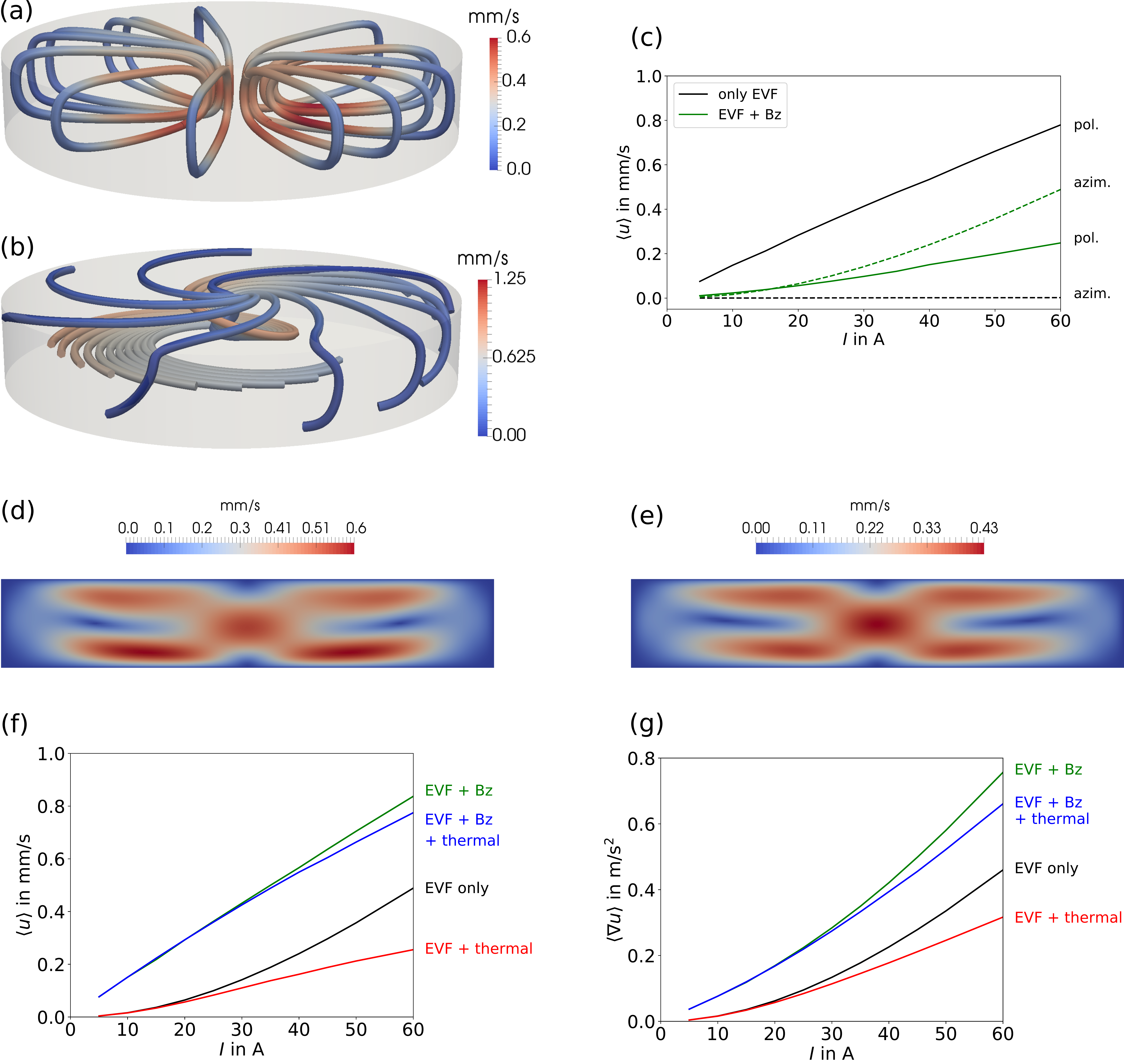}
\caption{Streamlines and velocity without (a) and with the Earth
  magnetic field (b). Volume averaged mean velocities of the azimuthal
  and poloidal flow for both cases (c). Velocity on a vertical plane
  for symmetric current supply without temperature (d) and with a
  negative temperature gradient of 0.7\,K (e). Volume averaged mean
  velocity (f) and mean velocity gradient evaluated at the top surface (g) of electro-vortex flow alone, with an additional Earth
  magnetic field (Bz) and with a stabilising temperature gradient. 
  $I=40$\,A.}
\label{f5}
\end{figure}

\begin{figure}[tbh!]
\centering
\includegraphics[width=\textwidth]{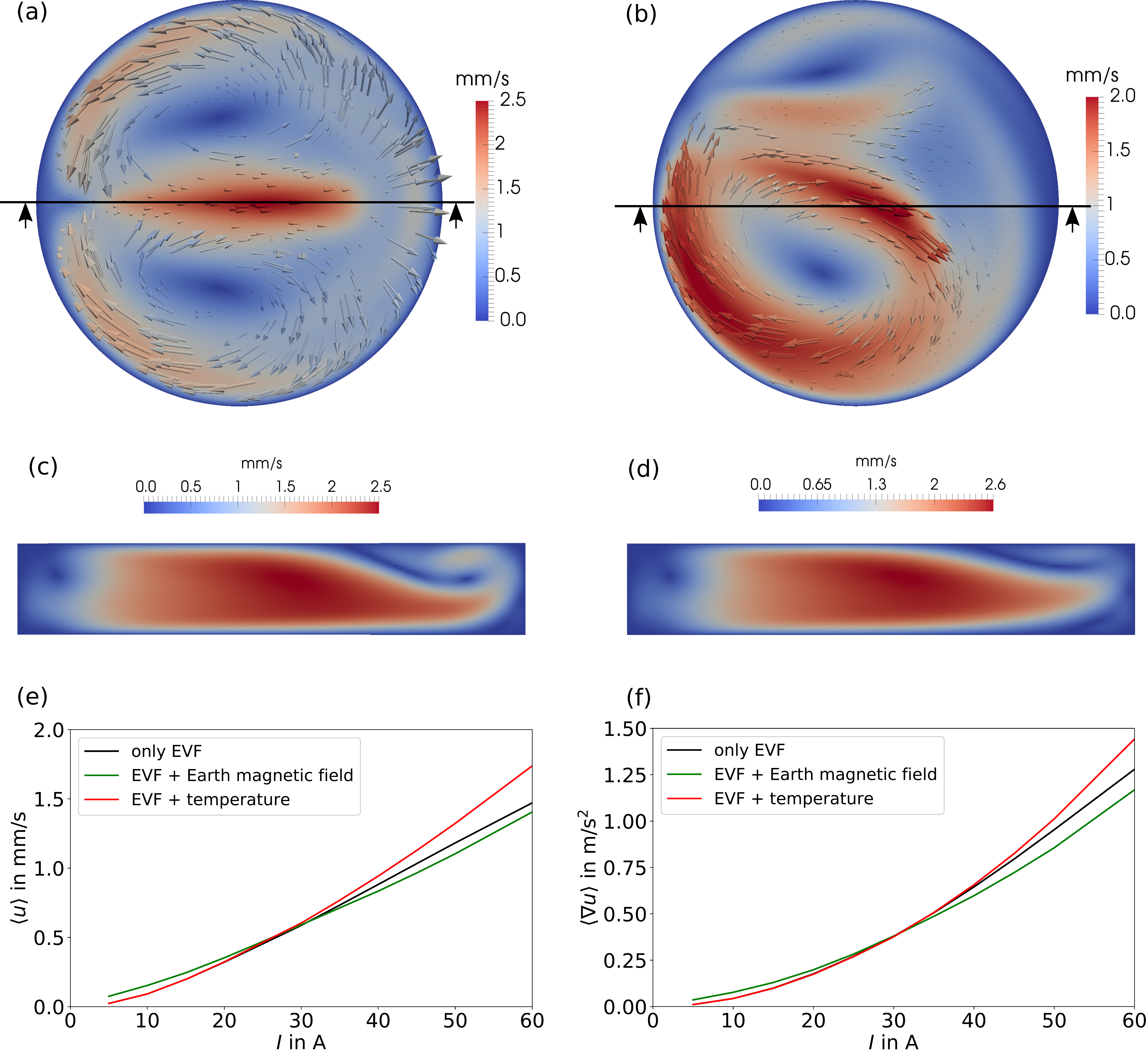}
\caption{Electro-vortex flow for a lateral supply wire without (a) and
  with the Earth magnetic field (b). Flow in the cross section of the
  jet without (c) and with a stabilising thermal gradient (d). The
  current for (a)-(d) is 40\,A. Volume averaged mean velocity (e) and mean
  velocity gradient (f) for
  electro-vortex flow alone, with the additional Earth magnetic field 
  and with a stabilising temperature gradient.}
\label{f6}
\end{figure}

This section is arranged as follows: firstly, we compare the
influence of a symmetric and asymmetric current supply on pure
electro-vortex flow (fig.~\ref{f4}). Thereafter, we study the
influence of the Earth magnetic field and of thermal stratification on
both connection types (fig.~\ref{f5} and \ref{f6}). Further, we give
estimates of the flow velocity depending on the cell current. 

Fig.~\ref{f4}a illustrates the current path, streamlines and
velocities for a lateral supply line. Electro-vortex flow is simulated
alone; the applied current is 40\,A. The flow profile is essentially
horizontal forming two kidney-shaped vortices. The velocity reaches
\SI{2.5}{\milli\meter/\second}. The \emph{horizontal} jet (also shown in \ref{f4}c) is
uncommon for electro-vortex flow, but can easily be explained. As the
current flows mainly horizontally through the copper plate, it induces
a magnetic field in the fluid. This field points towards the observer
(in fig.~\ref{f4}a and c). The current in the liquid metal flows
upwards (vertically) and interacts with the induced
field. Consequently, the
Lorentz force points to the right and drives the observed flow in
``prolongation'' of the current supply. For similar experiments, see
\cite{Dementev1992,Kolesnichenko2005}. 

Fig.~\ref{f4}b shows the flow for a symmetric current supply,
again at 40\,A. A typical poloidal flow develops as it was often
observed experimentally
\cite{Woods1971,Butsenieks1976,Boyarevich1977,Bojarevics1989,Zhilin1986,Raebiger2014}. Similar
flow structures are very well known from vacuum arc remelting and
electro-slag remelting
\cite{Atthey1980,Davidson1994,Davidson1999,Davidson2000,Kharicha2008,Shatrov2012,Kazak2012,Kharicha2015}. However,
depending on the exact geometry, the direction of the flow might be
reversed
\cite{Kazak2010,Kazak2011,Kazak2013,Semko2014}. In our simulation, the
velocities reach \SI{0.6}{\milli\meter/\second} for the symmetrical setup. This is
only 25\,\% of the flow velocity observed for a lateral current
supply. Due to the shallow liquid metal layer, a poloidal flow will
dissipate strongly in the boundary layer.

The simulated velocities are not directly comparable to the
experiment\cite{Kelley2014}. The latter was additionally heated from below (vertical
temperature difference of approximately $\Delta T=10$\,K). As shown
numerically by Beltr\'an, the experimentally observed flow is mainly
caused by Rayleigh-B\'enard convection. Also he used the linear
expansion coefficient \cite{Kelley2014,Beltran2016} which is three 
times smaller than the volumetric one  
\cite{Ashour2017a}, his velocity profile and magnitude (\SI{3}{\milli\meter/\second})
matches very well to the experimental results (compare fig.~9 in
\cite{Beltran2016} and fig.~4 in \cite{Kelley2014}). Electro-vortex
flow will generally lead to velocities one order of magnitude smaller
(Kelley and Sadoway \cite{Kelley2014} used currents of $23.3$\,A at
most; our results are for $40$\,A). However, 
electro-vortex flow will surely influence the flow structure and may
explain the increase in order of the flow which was observed
experimentally.

In the next step we focus on the symmetric current supply (with the poloidal
flow) only, and analyse the influence of 
a vertical magnetic background field. When we add the magnetic field
of the Earth (measured in Dresden as $\bi B=(15\cdot\bi e_x, 5\cdot\bi e_y, 36 \cdot\bi
e_z)$\,$\mu$T), the original poloidal flow (fig.~\ref{f5}a) becomes
strongly helical (fig.~\ref{f5}b). The appearance of such azimuthal
swirl flow is well known from experiments \cite{Woods1971,
  Bojarevics1983, Ashour2017a} and can be easily explained. Radial
cell currents and a vertical magnetic background field lead to
azimuthal Lorentz forces
\cite{Boyarevich1977,Bojarevics1983,Davidson1999}.
Compared to a recent experiment by Ashour et al.
\cite{Ashour2017a} with a point electrode on the top, we observe
considerably stronger swirl (compare fig.~\ref{f5}b with fig.~5 in
\cite{Ashour2017a}). We attribute this difference to the location of
the azimuthal forcing. Here, the force is well distributed in the whole
volume; in \cite{Ashour2017a} it is concentrated only in the centre of
the liquid metal ``sheet''. We suppose the distributed azimuthal
Lorentz force to better suppress the poloidal flow by forcing the
streamlines into a dissipative Ekman layer
\cite{Davidson1999}. Fig.~\ref{f5}c shows the volume averaged mean
velocity of the poloidal and azimuthal flow -- with and without the
Earth magnetic field. If we add a vertical field, azimuthal swirl
appears (compare the dashed curve). At the same time, the poloidal flow
is strongly reduced (by a factor of 1/2). This fits nicely to
Davidsons ``poloidal suppression'' model \cite{Davidson1999}. This is
remarkable, because simulations with a point electrode (see
\cite{Ashour2017a}) did not show such a strong suppression.

Keeping the symmetric current supply, we now focus on the influence of
the thermal stratification. During operation of an LMB, the cathode will be heated 
from above; the thermal  stratification will be stable. At first
glance, this configuration is similar to arc remelting. There, an
electric arc heats the melt from above. However, the bath is cooled
rather from the side than from below which leads to strong thermally driven
flow \cite{Davidson2000a}, but we have a stable thermal stratification instead. Based on the
heat conduction model described in section \ref{s:model} we
apply a vertical temperature gradient of $\Delta T = 0.7$\,K (at 40\,A). The
stable thermal stratification dampens the electro-vortex flow
(compare fig.~\ref{f5}d and e). While the general flow structure does
not change, especially the velocity near the bottom wall decreases by
a factor of 2/3. This result cannot be compared to the experiment, as
Kelley and Sadoway heated from below (and we from above). A temperature
gradient as in the experiment is not expected to appear during
``normal'' operation of an LMB; however, an additional heating or
cooling for mass transfer enhancement (as proposed in
\cite{Bradwell2011a,Bradwell2015}) can easily lead to similar
configurations.

Using a thermal diffusivity of $\alpha=6\cdot10^{-6}$\,m$^2$/s, a mass
diffusivity of $D=1.2\cdot10^{-8}$\,m$^2$/s \cite{Kelley2017}, a typical velocity scale
of $u=$\SI{1}{mm/s} and the height of the liquid metal $\Delta h=16$\,mm, 
we find a thermal Peclet number of $\text{Pe}_\text{th}=u\Delta h/\alpha\approx 5$
and a concentration Peclet number of $\text{Pe}_\text{c}=u\Delta h/D\approx 6\,000$ \cite{Deen1998}. 
Obviously, convection dominates mass transfer. We use thererfore 
two quantities to estimate mixing in the cathode: the
volume averaged velocity as global measure, and the mean velocity
gradient at the foam-cathode interface as local one. Fig.~\ref{f5}f and g
show both quantities for 
electro-vortex flow alone, with the Earth magnetic field ("Bz") and
with a stabilising thermal gradient. The azimuthal flow, caused by the
Earth magnetic field, yields the highest velocities. A vertical
temperature gradient does barely influence the horizontal
flow. The poloidal electro-vortex flow (``EVF alone'') is considerably
slower -- it is strongly dissipated at the bottom wall. The vertical
temperature gradient effectively breaks the downwards flow. Interestingly,
a strong flow in the volume leads also to strong velocity gradients at
the interface.

We now consider the lateral current supply, and study again the
influence of temperature and the Earth magnetic field. The prevailing
horizontal flow is hardly influenced by a stabilising vertical
temperature gradient. The flow structure changes only slightly; the
velocities with and without temperature gradient are almost the same
(compare fig.~\ref{f6}c and d). Taking into account the Earth magnetic field changes
the flow much more (compare fig.~\ref{f6}a and b). The horizontal current
and vertical magnetic background field generate a Lorentz force which
deflects the jet in clockwise direction. Presumably the stronger
dissipation in the boundary layers decreases the velocity
slightly. Most importantly, the Earth magnetic field does not lead to
swirl flow in this configuration -- the jet is only
deflected. Fig.~\ref{f6}e and f show the mean velocity 
and the mean velocity gradient for
pure electro-vortex flow, with the Earth magnetic field and with the
stabilising temperature gradient. The differences are only marginal.

\section{Summary \& outlook}
We have shown, how electro-vortex flow (EVF) has the potential to enhance mass
transfer in liquid metal batteries (LMBs). In a first step we
discussed why such mass transfer enhancement is important. 
Considering the high concentration Peclet number (in the order of $10^3$), we pointed out
that convection (and not diffusion) will dominate mass transfer.
We further emphasised
that mostly (but not only) mixing of the positive electrode during discharge is
highly beneficial. 
We studied the flow structure and magnitude of EVF
numerically. Moreover, we discussed the influence of stray magnetic
fields, the connection of the supply lines and a stable thermal
stratification on electro-vortex flow.

A lateral current supply to the cathode will generate a
\emph{horizontal} flow. In contrast, a centrical current supply below
the cathode will induce a \emph{vertical} jet. Looking only on this 
flow-direction, we would expect a vertical flow to be better suited for enhancing
mass transfer. It will remove reaction products directly from the
cathode-electrolyte interface. However, the vertical (or better:
poloidal) flow has three disadvantages: (1) it's mean velocity is much
smaller compared to the horizontal flow, (2) it is dampened by the
stable thermal stratification and (3) it will turn to a swirling
flow under presence of the Earth magnetic field. In contrast, the
horizontal jet will not be dampened considerably by a thermal
stratification nor be strongly influenced by the Earth magnetic
field. We believe therefore the lateral supply line to be better
suited for enhancing mass transfer. Concerning the swirl flow we could
(at least partially) confirm Davidsons model of poloidal suppression. 

Our models are strongly simplified: we ignore induced currents and
magnetic fields; the simulation of thermal
convection and EVF is decoupled. A next step would be therefore the
development of a fully coupled EVF-thermal convection model as well as
it's coupling with a real mass transfer (e.g. Li in Bi) model. Of course,
velocity and concentration measurements in a real 3-layer LMB would
be a large step forward. Performing Kelley \& Sadoway's experiment with an
inverse temperature gradient could allow a further experimental 
study of the interaction between EVF and thermal convection. Such an
experiment should preferably conducted at room temperature to
ensure well defined boundary conditions for temperature.

\section*{Acknowledgements}
This work was supported by the Deutsche Forschungsgemeinschaft (DFG,
German Research Foundation) by award number
338560565 as well as the Helmholtz-Gemeinschaft Deutscher
Forschungs\-zentren (HGF) in frame of the Helmholtz Alliance
``Liquid metal technologies'' (LIMTECH). The computations were
performed on the Bull HPC-Cluster ``Taurus'' at the Centre for
Information Services and High Performance Computing (ZIH) at TU
Dresden and on the cluster ``Hydra'' at Helmholtz-Zentrum Dresden --
Rossendorf. Fruitful discussions with V. Bojarevics, P. Davidson,
D. Kelley, F. Stefani and T. Vogt on several aspects of electro-vortex
flow and thermal convection are gratefully acknowledged. N. Weber
thanks Henrik Schulz for the HPC support. 

\section*{References}
\bibliography{literature}

\end{document}